\documentclass[sigconf]{acmart}
\usepackage{subcaption}
\usepackage{url}
\usepackage{enumitem}
\usepackage{graphicx}
\usepackage{caption}
\usepackage{amsmath}
\usepackage{tabu}
\usepackage{multirow}
\usepackage{hyperref}
\usepackage{adjustbox}
\usepackage{mathptmx}
\usepackage{amsfonts}

\usepackage{array} 
\usepackage{makecell} 
\usepackage{fontawesome}
\usepackage{tikz}
\newcommand*\circled[1]{\tikz[baseline=(char.base)]{
            \node[shape=circle,draw,inner sep=0.75pt, text=white,fill=black] (char) {#1};}}

\usepackage{algpseudocode}
\usepackage{pgfplots}
\pgfplotsset{compat=1.12}
\usepackage{soul}
\usepackage{footnote}
\usepackage{lipsum}
\usepackage[ruled,vlined, linesnumbered]{algorithm2e}
\usepackage{subcaption}
\usepackage{listings}
\def\BibTeX{{\rm B\kern-.05em{\sc i\kern-.025em b}\kern-.08em
    T\kern-.1667em\lower.7ex\hbox{E}\kern-.125emX}}
    
\newcommand\YAMLcolonstyle{\color{red}\mdseries}
\newcommand\YAMLkeystyle{\color{black}\bfseries}
\newcommand\YAMLvaluestyle{\color{blue}\mdseries}
\makeatletter
\newcommand\language@yaml{yaml}
\expandafter\expandafter\expandafter\lstdefinelanguage
\expandafter{\language@yaml}
{
  keywords={true,false,null,y,n},
  keywordstyle=\color{darkgray}\bfseries,
  basicstyle=\YAMLkeystyle,                                 
  sensitive=false,
  comment=[l]{\#},
  morecomment=[s]{/*}{*/},
  commentstyle=\color{purple}\ttfamily,
  stringstyle=\YAMLvaluestyle\ttfamily,
  moredelim=[l][\color{orange}]{\&},
  moredelim=[l][\color{magenta}]{*},
  moredelim=**[il][\YAMLcolonstyle{:}\YAMLvaluestyle]{:},   
  morestring=[b]',
  morestring=[b]",
  literate =    {---}{{\ProcessThreeDashes}}3
                {>}{{\textcolor{red}\textgreater}}1     
                {|}{{\textcolor{red}\textbar}}1 
                {\ -\ }{{\mdseries\ -\ }}3,
}

\lst@AddToHook{EveryLine}{\ifx\lst@language\language@yaml\YAMLkeystyle\fi}
\makeatother

\newcommand\ProcessThreeDashes{\llap{\color{cyan}\mdseries-{-}-}}

\newcolumntype{P}[1]{>{\centering\arraybackslash}p{#1}}

\usepackage[nolist]{acronym}

\usepackage{float}

\begin{acronym}

\acro{faas}[FaaS]{Function-as-a-Service}
\acro{iaas}[IaaS]{Infrastructure-as-a-Service}
\acro{paas}[PaaS]{Platform-as-a-Service}
\acro{baas}[BaaS]{Backend-as-a-Service}
\acro{hpc}[HPC]{High-Performance Computing}
\acro{slo}[SLO]{Service-Level Objective}
\acro{sla}[SLA]{Service-Level Agreements}
\acro{rtt}[RTT]{Round Trip Time}
\acro{vu}[VU]{Virtual User}
\acro{rt}[RT]{Response Time}
\acro{mms}[MMS]{Minimum Memory Sum}
\acro{moc}[MOC]{Minimum Overall Cost}
\acro{csp}[CSP]{Cloud Service Provider}

\end{acronym}

\AtBeginDocument{%
  \providecommand\BibTeX{{%
    \normalfont B\kern-0.5em{\scshape i\kern-0.25em b}\kern-0.8em\TeX}}}

\copyrightyear{2021} 
\acmYear{2021} 
\acmPrice{15.00}

\begin{document}

\title{Estimating the Capacities of Function-as-a-Service Functions}
\subtitle{*This is the preprint version of the accepted paper at CloudAM'21 workshop (UCC)}

\author{Anshul Jindal}
\orcid{0000-0002-7773-5342}
\affiliation{%
  \institution{Technical University of Munich}
  \streetaddress{Boltzmannstraße 3}
  \city{Garching (near Munich)}
  \state{Bavaria}
  \country{Germany}
  \postcode{85748}
}
\email{anshul.jindal@tum.de}

\author{Mohak Chadha}
\affiliation{%
  \institution{Technical University of Munich}
  \streetaddress{Boltzmannstraße 3}
  \city{Garching (near Munich)}
  \state{Bavaria}
  \country{Germany}
  \postcode{85748}
}
\email{mohak.chadha@tum.de}

\author{Shajulin Benedict}
\affiliation{%
  \institution{Indian Institute of Information Technology}
  \city{Kottayam}
  \state{Kerala}
  \country{India}
}
\email{shajulin@iiitkottayam.ac.in}

\author{Michael Gerndt}
\affiliation{%
  \institution{Technical University of Munich}
  \streetaddress{Boltzmannstraße 3}
  \city{Garching (near Munich)}
  \state{Bavaria}
  \country{Germany}
  \postcode{85748}
}
\email{gerndt@in.tum.de}

\renewcommand{\shortauthors}{A. Jindal et al.}

\begin{abstract}
Serverless computing is a cloud computing paradigm that allows developers to focus exclusively on business logic as cloud service providers manage resource management tasks. Serverless applications follow this model, where the application is decomposed into a set of fine-grained Function-as-a-Service (FaaS) functions. However, the obscurities of the underlying system infrastructure and dependencies between FaaS functions within the application pose a challenge for estimating the performance of FaaS functions. To characterize the performance of a FaaS function that is relevant for the user, we define Function Capacity (FC) as the maximal number of concurrent invocations the function can serve in a time without violating the \ac{slo}.

The paper addresses the challenge of quantifying the FC individually for each FaaS function within a serverless application. This challenge is addressed by sandboxing a FaaS function and building its performance model. To this end, we develop \textit{FnCapacitor} \textit{--} an end-to-end automated Function Capacity estimation tool. We demonstrate the functioning of our tool on Google Cloud Functions (GCF) and AWS Lambda. \textit{FnCapacitor} estimates the FCs on different deployment configurations (allocated memory \& maximum function instances) by conducting time-framed load tests and building various models using statistical: linear, ridge, and polynomial regression, and Deep Neural Network (DNN) methods on the acquired performance data. Our evaluation of different FaaS functions shows relatively accurate predictions with an accuracy greater than 75\% using DNN for both cloud providers.
\end{abstract}


\maketitle
\pagestyle{empty}

\section{Introduction}
\label{sec:intro}
With the advent of Amazon Web Services (AWS) Lambda in 2014, serverless computing has gained popularity and more adoption in different application domains such as machine learning~\cite{serverlessfl, carreira2019cirrus}, linear algebra computation~\cite{shankar2018numpywren, chard2019serverless}, and map/reduce-style jobs~\cite{jonas2017occupy}. Function-as-a-Service (FaaS), a key enabler of serverless computing, allows a traditional application to be decomposed into fine-grained functions that are executed in response to event triggers or HTTP requests on a FaaS platform~\cite{riseserverless}. The FaaS platform is responsible for deploying and providing resources to the FaaS functions.

Since serverless computing environments abstract the underlying system infrastructure configurations away from the users, most of the public cloud providers in their FaaS offerings allow users to configure only a small set of configuration parameters: memory allocation and the maximum number of function instances, also called as concurrency~\cite{AWSLambda, GoogleCloudFunctions:online, Azure}. Moreover, cloud providers speedup function execution when a higher memory is configured~\cite{perfcompute}. Furthermore, the heterogeneity in the underlying nodes can lead to variations in the execution time of the FaaS functions~\cite{arch-specific, courier}. Therefore, estimating the maximum number of requests the deployed functions can handle such that the response times adhere to the SLOs constraints poses a set of challenges.  Towards this, we define the maximal number of concurrent invocations that a FaaS function can serve within a time interval without violating the SLOs when deployed with particular memory configuration and fixed maximum function instances as the \textit{Function Capacity (FC)}. In this paper, we consider the $95^{th}$ percentile execution time of a FaaS function as the SLO. 

\begin{figure*}[t]
\centering
\begin{subfigure}{0.24\textwidth}
    \centering
        \includegraphics[width=1\linewidth]{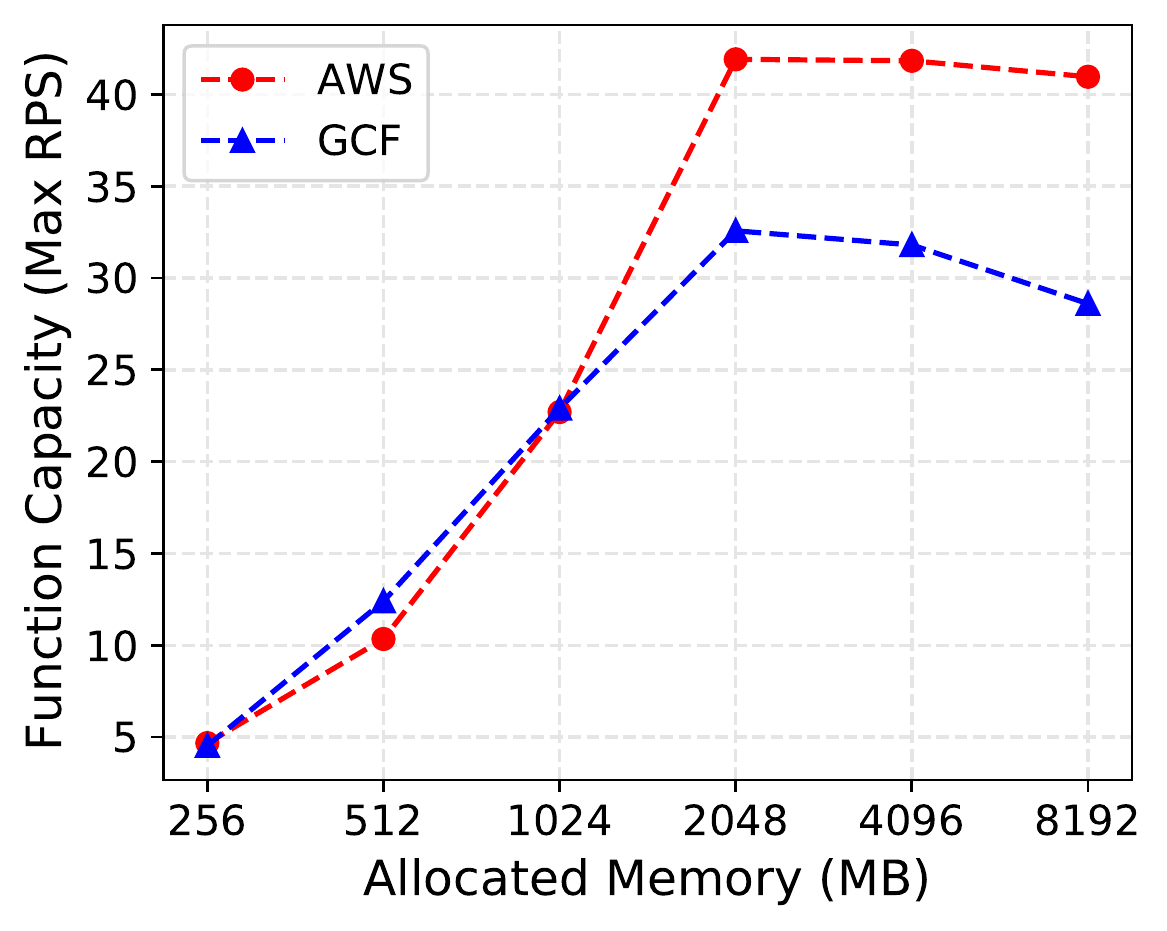}
        \caption{Effect of memory on the FC with fixed function concurrency of 100.}
        \label{fig:mem_capacity}    
\end{subfigure}\hspace{2pt}
\begin{subfigure}{0.24\textwidth}
    \centering
        \includegraphics[width=1\linewidth]{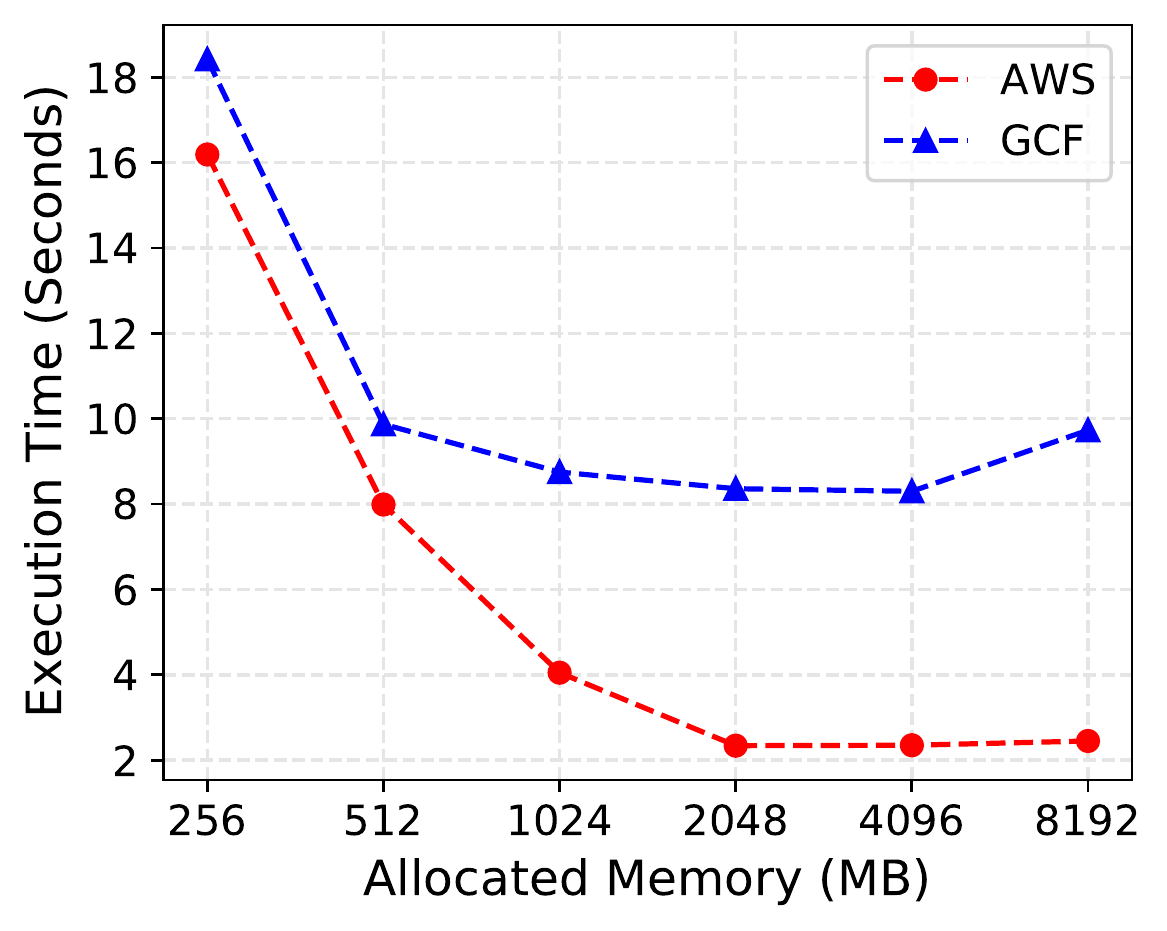}
        \caption{Effect of memory on the with fixed function concurrency of 100.  }
        \label{fig:runtime_capacity}    
\end{subfigure}\hspace{2pt}
\begin{subfigure}{0.24\textwidth}
    \centering
        \includegraphics[width=1.03\linewidth]{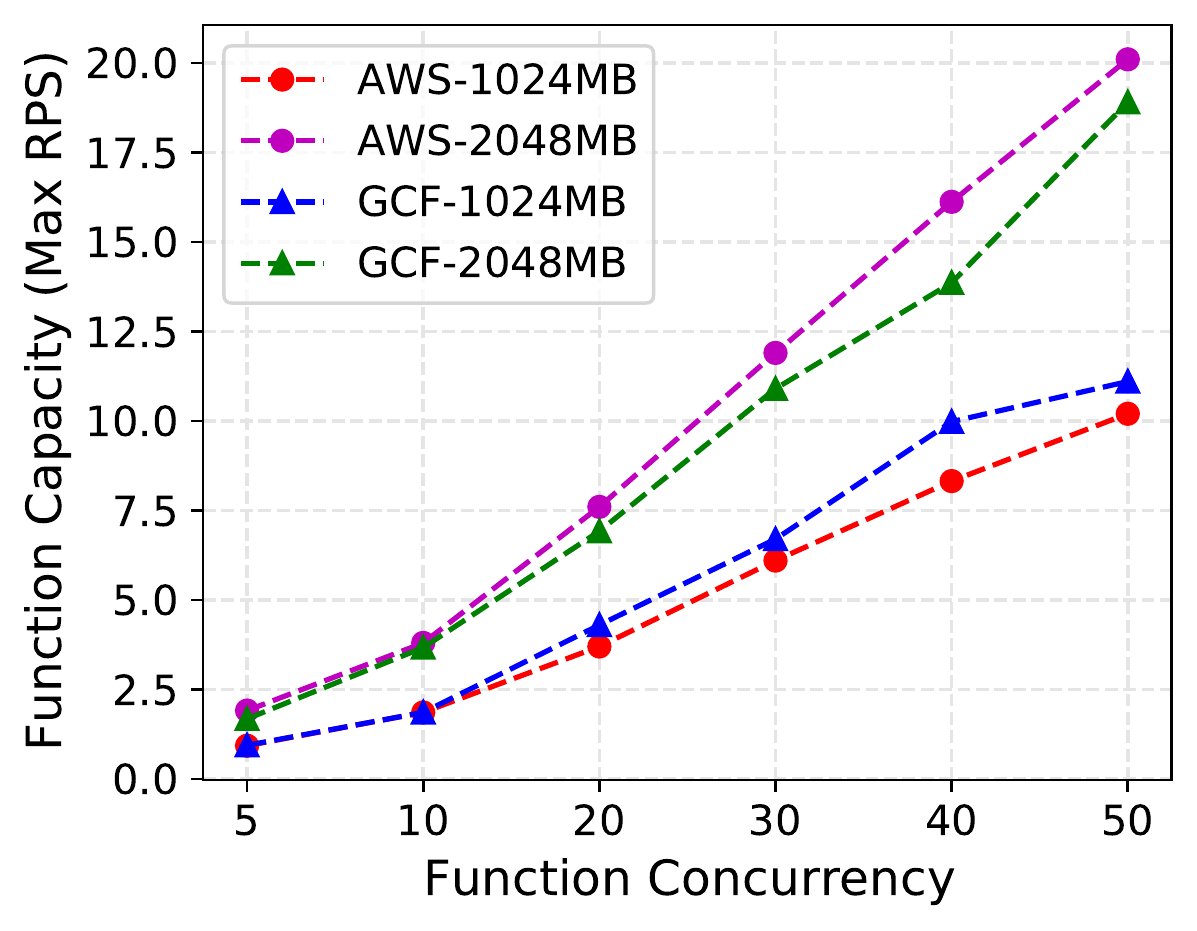}
        \caption{Function concurrency effect on the FC with fixed memory.}
        \label{fig:concurrency_capacity}    
\end{subfigure}
\caption{
AWS lambda and Google Cloud Function (GCF) FCs variation with memory configurations, and concurrency.}
\label{fig:multi_params}
\end{figure*}
To highlight the effects of various parameters configuration on the performance and the Function Capacities of FaaS functions, we deployed a compute-intensive (calculates prime numbers till $10000000$) serverless function written in Python on AWS Lambda and Google Cloud Function (GCF)~\cite{GoogleCloudFunctions:online}. We fixed the $95^{th}$ percentile execution time of the function to 20 seconds. Figure~\ref{fig:mem_capacity} shows how the Function Capacity (i.e., the maximum number of requests per second the function can handle) first increases with varying memory sizes upto a certain point (2048MB). After that, it becomes constant when the function concurrency is fixed to 100 for both the cloud providers. The same Figure~\ref{fig:mem_capacity} also shows the variation in FC with cloud providers. The variation in the system resources causes differences in performance between the identical function deployments for the same FaaS platforms. Figure~\ref{fig:runtime_capacity} shows the execution time of corresponding runs, and one can see that it decreases with the increase in memory, and after a point (2048MB), it also becomes constant.  Lastly, Figure~\ref{fig:concurrency_capacity} shows the linear increase in FC with the increase in function concurrency, keeping the memory fixed.  

The examples above highlight some factors that can affect the performance and the FCs. However, they are many other factors such as cold starts, I/O and network conditions, type of container runtimes, and co-location with other functions affecting the performance and FCs which the users are not aware of~\cite{wang2018peeking, closer20container}. Additionally, the dependencies between the functions within a serverless application can also affect the FCs. To this end, we develop \textbf{FnCapacitor}, a tool that can estimate the FCs of the functions adhering to the given SLOs, the specified memory configurations, and function concurrency. Our key contributions are as follows: 
\begin{itemize}
        \item We develop and present a novel python-based tool called \textit{FnCapacitor} for automatic estimation of Function Capacities of FaaS functions within a serverless application~(\S\ref{methodology}). 
        
        \item As part of \textit{FnCapacitor}, we present a functions sandboxing method which can be used to sandbox individual functions from the overall serverless application (\S\ref{sec:sandboxing}). 
        
        \item  We showcase the effect of different deployment configurations : memory allocation (\S\ref{sec:mem_effect_qos}) and function concurrency (\S\ref{sec:conc_effect_capacity}), on the Function Capacity for different FaaS functions within a sample application. To the best of our knowledge, none of the previous works involving FaaS~\cite{shahrad2019architectural, lee2018evaluation, wang2018peeking, fdn, 9155363, eismann2021sizeless} considered function concurrency parameter.    
        
        \item Although our approach is generic and \textit{FnCapacitor} can be easily extended to support other commercial and open-source FaaS platforms, we demonstrate the  functioning of \textit{FnCapacitor} with Google Cloud Functions (GCF) and AWS Lambda~(\S\ref{results}) on a sample serverless application consisting of \texttt{8} functions. We further present the performance results of the formulated FC estimation models on each of them (\S\ref{sec:capacity_prediction}). 
        
        \item We open source the collected data and developed tool for further research\footnote{\url{https://github.com/ansjin/faas_capacitor}}.
\end{itemize}

\textbf{Paper Organization} Section~\ref{taxonomy} introduces the basic FaaS function invocation procedure and Function Capacity concept. In Section~\ref{methodology} the developed tool \textit{FnCapacitor} is described. Section~\ref{exp_config} describes the experimental setup and the hyper-parameters of the different algorithms. In Section~\ref{results}, our evaluation results are presented. Section~\ref{related_work} describes some of the prior works in this domain. Finally, Section~\ref{conclusion} concludes the paper.



\section{Taxonomy}
\label{taxonomy}

\subsection{FaaS function invocation procedure}
\begin{figure}[t]
    \centering
    \includegraphics[width=0.7\linewidth]{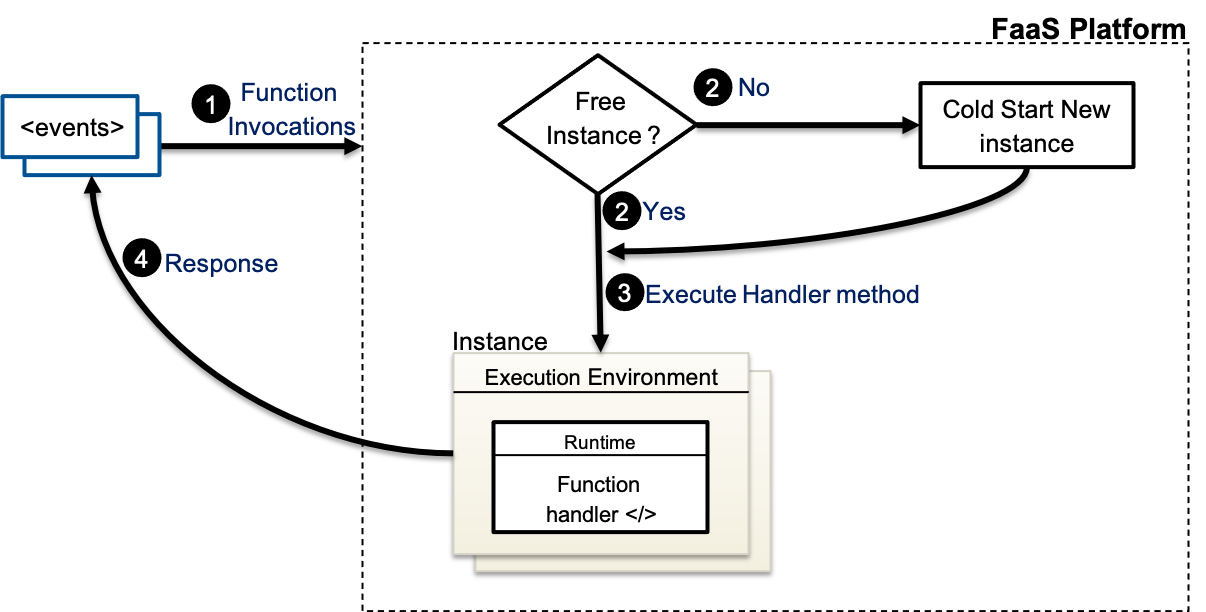}
    \caption{Typical FaaS function invocation procedure.}
    \label{fig:FaaS_invocation_procedure}
\end{figure}
Function-as-a-Service based \textit{function} is a piece of code containing a \textit{handler method} responsible for processing the \textit{events} that are passed to the function when invoked and these are executed within a \textit{FaaS platform}. FaaS based functions can be invoked by a user's HTTP request or another type of event created within the FaaS platform or the cloud infrastructure. These include changes to data in a database, files added to a storage system, or a new virtual machine instance is created. The FaaS platform is responsible for providing resources for function invocations and performs automatic scaling. This is done by creating an \textit{execution environment} which provides a secure and isolated runtime environment for the function. The functions can be written using various languages, and a language-specific environment called \textit{runtime} is created in the execution environment. The runtime relays invocation events, context information, and responses between the FaaS platform and the function. 

The first time the function is invoked, the FaaS platform creates an \textit{instance} of the function (execution environment) and runs its \textit{handler method} in it to process the event. When the handler exits or returns a response, it stays active and becomes available to handle other events. If the function is invoked again while the first event is being processed, the FaaS platform creates another \textit{instance}, and the two events are processed \textit{concurrently}. As more events come in, the FaaS platform routes them to available instances and creates new instances as needed. When the number of requests decreases, the FaaS platform stops unused instances to have free scaling capacity for other functions. FaaS platforms usually have an upper limit on how many maximum concurrent instances called \textit{function concurrency} can be created, such as 1000 for AWS lambda and 3000 for GCF. Figure~\ref{fig:FaaS_invocation_procedure} summarizes the overview of the typical FaaS function request procedure.

\subsection{Function Capacity (FC)}
We define the \textit{Function Capacity (FC)} as the maximal number of concurrent invocations that a FaaS function when deployed on a FaaS platform with a certain memory configuration and fixed maximum function instances, can serve within a time interval without violating the SLOs when deployed. Ideally, if an instance $i$ serves $n_i^t$ number of invocations within a time interval $t$ and $C$ is the function concurrency for the FaaS platform, then the $FC(f)$, where $f$ is any function, can be calculated by the equation~\eqref{eq:line}. 

\begin{equation}
FC (f) = n_i^t \times C \label{eq:line}
\end{equation}

However, in practice, there are many other factors such as cold starts, I/O and network conditions, type of container runtimes, and co-location with other functions affecting the performance and FCs of FaaS functions~\cite{wang2018peeking}. Therefore, we follow a modeling approach to estimate the capacities of the FaaS functions. In this paper, we consider the $95^{th}$ percentile execution time of a FaaS function as the SLO.

\section{FnCapacitor}
\label{methodology}
\begin{figure}[t]
    \centering
    \includegraphics[width=0.95\linewidth]{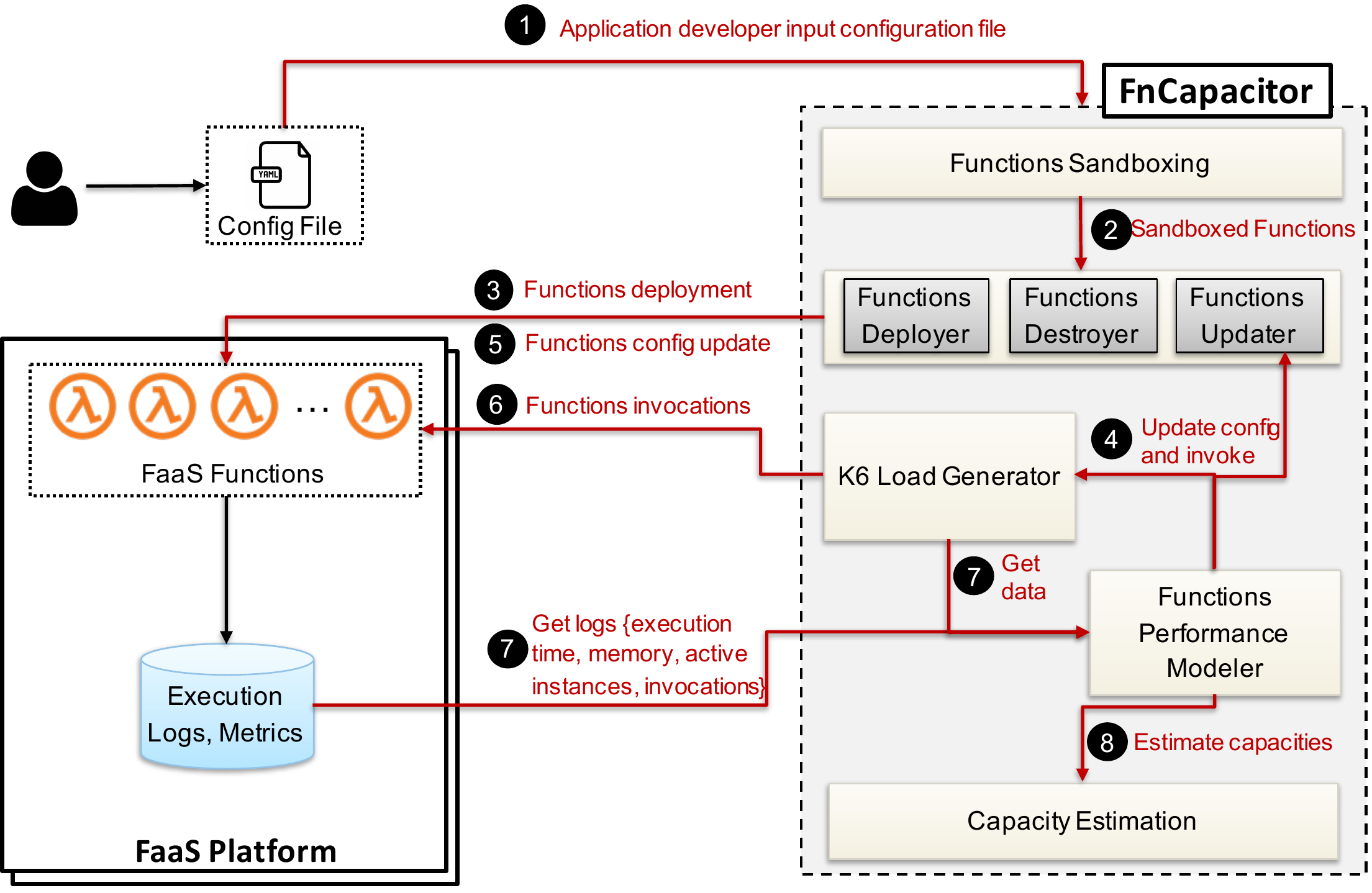}
    \caption{High-level architecture of the \textit{FnCapacitor}.}
    \label{fig:FaaSModeler}
\end{figure}

To estimate the capacities of the FaaS functions within a serverless application, we have developed \textit{FnCapacitor}, a python-based automated end-to-end function capacity estimation tool. Given the SLO requirements,  \textit{FnCapacitor} is responsible for estimating the FCs of FaaS functions at different deployment configurations (memory allocation and function concurrency). Figure~\ref{fig:FaaSModeler} provides an overview of the high-level architecture of \textit{FnCapacitor} and the interaction between its components.

\textit{FnCapacitor} takes a YAML file as input that specifies the initial FaaS platform configuration parameters (minimum memory allocation and functions timeout) for the function deployment, serverless application to be deployed, and configuration parameters for the load generator and the modeling (step \circled{1}). Since a serverless application consists of multiple functions and the performance of one function could affect the others depending upon it,  therefore in the next step, the individual functions from the given application are segregated (step \circled{2}). These sandboxed functions are then deployed on the FaaS platform with some initial configurations (step \circled{3}). After the deployment, \textit{FnCapacitor} repeatedly changes the functions configurations (steps \circled{4} - \circled{5}) and generates a user workload to the function's API endpoint (step \circled{6}) for collecting various monitoring metrics data (\S\ref{sec:monitoring_metrics}). The collected metrics data is stored in InfluxDB, which is used for creating the function performance models (step \circled{7}). The created function models are then used for estimating the Function Capacities for different deployment configurations (step \circled{8}).

We now describe the sub-components of \textit{FnCapacitor}: 

\subsection{Functions Sandboxing}
\label{sec:sandboxing}
Usually, a serverless application consists of multiple functions, and the performance of one function could affect the others depending upon it. Therefore to measure the pure performance of the functions i.e., where their performance is not affected by others, we build this component for sandboxing individual functions through a mockup of their neighbors. It isolates each function and substitutes its direct neighbors with dummy functions accepting the requests and sending the responses in the same format, but without any additional processing allowing to measure the pure performance of only that function and build models using this data.

Firstly, this component for each function within the serverless application replaces the calls to other functions with calls to a \textit{proxy-function}. This  \textit{proxy-function} serves as an intermediator between the sandboxed function and other functions and takes the originally called function names from the sandboxed function and the input payload to them as the input. This allows every invocation to other functions to go through this \textit{proxy-function} and this dummy \textit{proxy-function} then will invoke the following functions based on the input received. At the same time, copies of these requests and responses are stored in the FnCapacitor's MongoDB database for creating function mockups. It is to be noted that, \ac{baas} services such as database, storage, queues, etc., are out of the scope of this work for sandboxing as it is assumed that these \ac{baas} services provides high scalability and serve the user requests within the defined \acp{slo}. Following this, each function receives its own sandboxed deployment where mockup functions replace the direct neighbors. These mockup functions will respond with the response stored in MongoDB. As a result, the time taken by the dependent functions to respond becomes negligible and therefore allows to build a pure performance model of the functions.

\subsection{Function Deployer, Destroyer and Updater}
As the name suggests, it is responsible for deploying, destroying, and updating the functions to be modeled on the GCF and AWS Lambda using the serverless framework~\cite{serverlessFramework}. For deployment and updating, different configuration parameters: memory configurations and function concurrency are taken into account. The deployed functions are scaled automatically by adding multiple function instances by the FaaS platform. This component automatically deletes the functions when the test is finished to free up the resources.

\subsection{Load Generator}
The Load Generator is implemented using a load testing tool - \textit{k6}. It uses a script for running the tests where the function endpoint and request parameters are specified. As part of each test,  the number of requests per second (RPS) generated by \textit{k6} is varied and depends on the number of Virtual Users (VUs) and the time taken by each request to complete. VUs are the entities in \textit{k6} that execute the test and make HTTP(s) requests. The load generation metrics from \textit{k6} and FaaS platform monitoring metrics data from the cloud providers are exported and stored inside an InfluxDB instance. 


\subsection{Performance Modeler \& Capacity Estimation}
\label{sec:modeler}
This is the main component of \textit{FnCapacitor} and is responsible for analyzing the correlation between the different monitoring metrics (\S\ref{sec:monitoring_metrics}) and the deployment configurations. It uses the collected data stored in InfluxDB to create models of the functions and estimate their Function Capacities. Modeling approaches used in this work are categorized under two categories: 
\begin{itemize}
     \item \textbf{Statistical Approaches}:  We consider linear, polynomial, ridge, and random forest regression for modeling the relationship.
     \item \textbf{Deep Neural Networks (DNNs)}: DNNs are designed to solve complex problems by building relationships among multiple dependent and independent variables. As a result, we use them for formulating the models. 
\end{itemize}

The collected data from InfluxDB is pre-processed by removing outliers and dividing the data into training and test set. Following this, different models, i.e., statistical and DNNs, are trained on the training data set. Due to the training data being sparse, k-fold cross-validation (in our case k=6) is followed for training the model~\cite{yadav2016analysis}. 

Lastly, Function Capacities are estimated on the new or test dataset using the trained models. The prediction accuracy is computed using the $R^2$ score value from the actual and predicted Function Capacities for the test dataset.

\section{Experimental Configuration}
\label{exp_config}
In this work, we have fixed the total duration of a test to \texttt{30} minutes for the deployed serverless application. A test consists of the memory allocation configurations: \texttt{<256MiB, 512MiB, 1GiB, 2GiB and 4GiB>} and function concurrency: \texttt{<10, 20, 30, 40 and 50>}.  AWS functions are deployed in the europe-central1 region, and GCF functions are deployed in the eu-west3 region. The number of VUs during the load generation are varied from 5 to 500 depending on the number of requests the functions can serve. As a result, for each function, $5$ (\text{total memory configurations}) $\times$ $5$ (\text{function concurrency configurations}) $= 25$ tests were conducted. 


We partition the collected data into training and test set (33\% of the total data). We used a part of the training data set as a validation set for selecting the hyperparameters of the different models. We select the hyperparameters through an exhaustive grid search. For the DNN model, we use a 12-layer fully connected neural network architecture, with each layer having 64 units. We use the Rectified Linear Unit (ReLU) as activation function~\cite{imagenet}. We use mean absolute error as the loss function for training the DNN model with Adam as the optimizer.

\subsection{Application used for evaluation}
\label{microbenchmarks}
\begin{figure} [t]
    \centering
    \includegraphics[width=1\columnwidth]{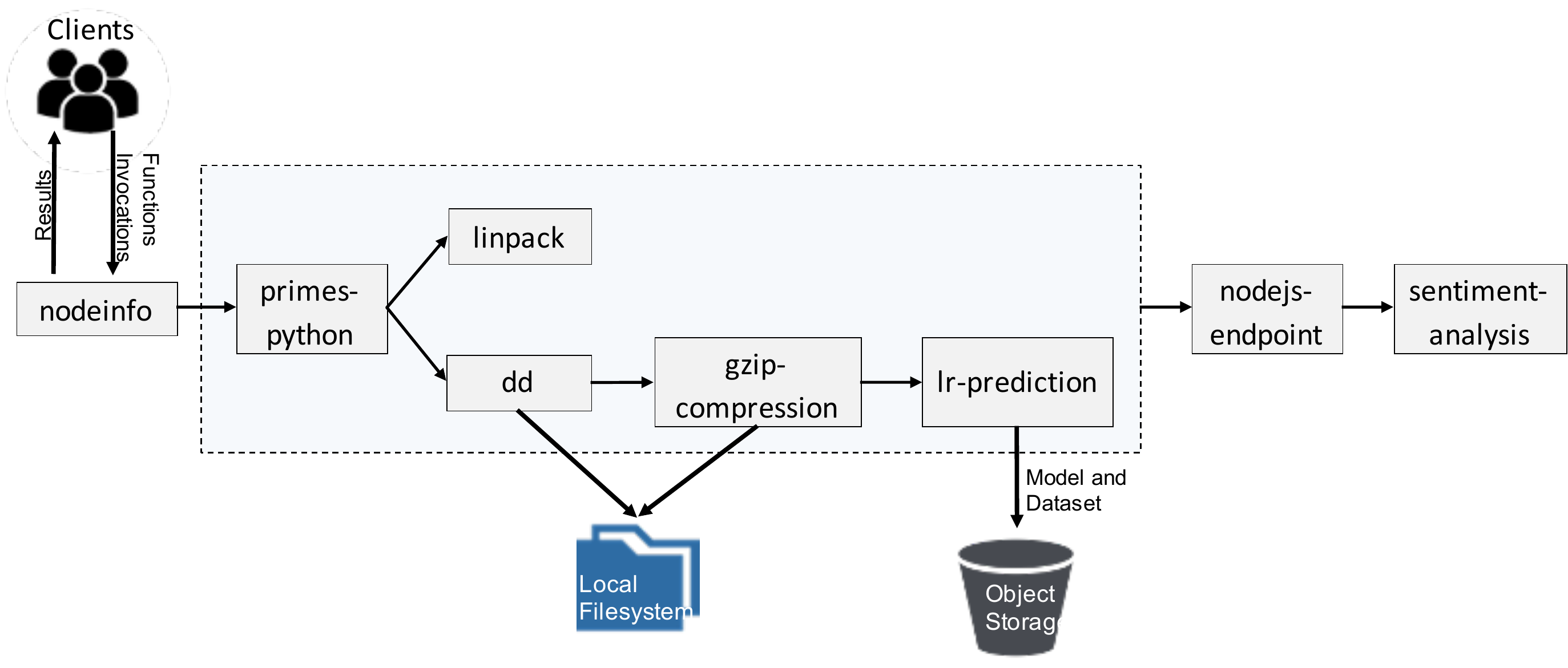}
    \caption{High level workflow of the evaluated application.}
    \label{fig:benchmark-application}
\end{figure}

To investigate the performance of each deployment configuration, we used a subset of the benchmarks provided with the FaaSProfiler~\cite{shahrad2019architectural}. We created an application shown in Figure~\ref{fig:benchmark-application} for our use case. The functions used as part of this application are summarized in Table~\ref{functions_used}. 
\begin{table*}[t]%
\centering
\caption{ Functions comprising the evaluated application. \label{functions_used}}%
\begin{tabular}{|P{2.5cm}|m{11.5cm}|P{2cm}|}
 \hline
\multirow{1}{*}{\textbf{Functions}}  & \multirow{1}{*}{\textbf{Description}} & \textbf{Runtime}\\
 \hline
 \multirow{1}{*}{nodeinfo}   & Gives basic characteristics of node like CPU count, architecture, uptime. & \multirow{1}{*}{Node.js 14} \\
  \hline
primes-python  & Calculates prime numbers till $100000$. & Python3   \\
 \hline
\multirow{1}{*}{linpack} &  It solves a
dense linear system of equations in double precision and returns the results in GFlops. Problem size (number of equations) is fixed to 100. 
& \multirow{1}{*}{Python3} \\
\hline
\multirow{1}{*}{dd} & It is based on Unix dd command-line utility for converting and copying files. 128bytes as a block size and 5 times conversion is used as parameters. & \multirow{1}{*}{Python3} \\
\hline
\multirow{1}{*}{gzip-compression} & It first creates a file filled with random numbers of size 1MB and then compresses it using gzip compression scheme. & \multirow{1}{*}{Python3} \\
 \hline
 \multirow{1}{*}{lr-prediction} & It first downloads a linear regression model trained on user reviews data from the storage bucket along with the test data and performs prediction on it. & \multirow{1}{*}{Python3} \\
 \hline
 \multirow{1}{*}{nodejs-endpoint} & It is a simple REST endpoint which returns the current time along with the time zone. & \multirow{1}{*}{Node.js 14} \\
 \hline
  \multirow{1}{*}{sentiment-analysis} & Analyzes the sentiment of a provided string using the Python TextBlob library & \multirow{1}{*}{Python3} \\
 \hline

\end{tabular}
\end{table*}

The application flow starts with the \texttt{nodeinfo} function, which exposes an HTTP endpoint and provides the user with basic information about the system such as hostname, underlying architecture, number of CPUs, etc. This then invokes the compute-intensive \texttt{primes-python} function to judge the built models' abilities on compute-intensive applications, and it further invokes \texttt{linpack} and \texttt{dd} asynchronously and waits for their response to come back. \texttt{dd} invokes \texttt{gzip-compression}. It further invokes \texttt{lr-prediction} in a sequence. \texttt{lr-prediction} queries the model and data from the google cloud storage (created in the google compute platform in the europe-west3 region, AWS lambda functions also use this storage bucket) and then performs prediction.  Once the response is available to \texttt{primes-python} function from both the invocations, it sends back the response to \texttt{nodeinfo} function, which in turn invokes \texttt{nodejs-endpoint} and \texttt{nodejs-endpoint} further invokes \texttt{sentiment-analysis} function. We use these eight heterogeneous functions to test the accuracy of the modeling on different types of functions.

\subsection{Monitoring Metrics}
\label{sec:monitoring_metrics}
We extracted following monitoring metrics from the GCF and AWS lambda\footnote{https://docs.aws.amazon.com/lambda/latest/dg/monitoring-metrics.html} with the data sampling rate as one minute:

\begin{itemize}
     \item \textit{concurrent\_instances}: Number of concurrent function instances. 
     \item \textit{invocations}: Number of times the function is executed. 
     \item \textit{execution\_duration}: Amount of time function code spends in processing an event.
     \item \textit{memory\_usage}: Function's  maximum  memory usage. 
     \item \textit{allocated\_memory}: Memory allocated to the function.
      \item \textit{function\_concurrency}: Maximum number of concurrent instances allowed for processing events. 
\end{itemize}

%





\section{Experimental Results}
\label{results}

In this section, we first describe the impact of heterogeneity in the memory allocations on function's \textit{execution\_duration} and \textit{concurrent\_instances}, and then the effect of \textit{function\_concurrency} on the Function Capacities for both the cloud providers and on different functions. Following this, we present results of FCs estimation using the different modeling approaches.

\subsection{Memory effect on function execution duration}
\label{sec:mem_effect_qos}


\begin{figure}[t]
    \centering
    \includegraphics[width=1\linewidth]{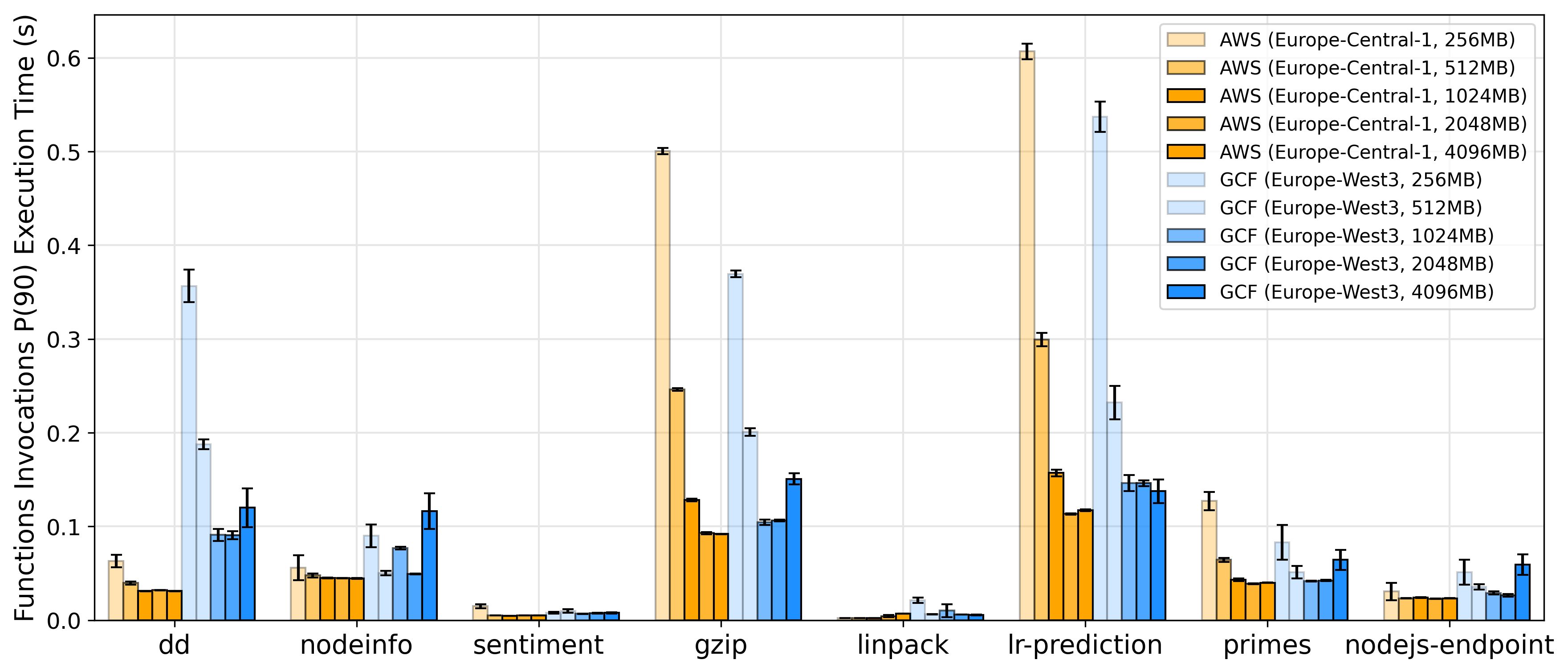}
    \caption{\textit{execution\_duration}s of the sandboxed functions when executed with a load of $50$ RPS with five different memory configurations and no limit on the \textit{function\_concurrency}. }
    \label{fig:execution_time_no_limit}
\end{figure}
Figure~\ref{fig:execution_time_no_limit} shows the \textit{execution\_duration}s of the load testing of $50$ invocations per second on the application when the functions are sandboxed and deployed with five different memory configurations on GCF and AWS lambda platforms. We observe the following: 

\textbf{Decrease in \textit{execution\_duration} with the increase in memory and becoming constant}: From Figure~\ref{fig:execution_time_no_limit}, we can see that for most of the functions and across two FaaS platforms, there is a general trend of decrease in \textit{execution\_duration} with the increase in memory, and after a certain point (2048MB memory configuration), either it is becoming constant (for all functions running on AWS, and \textit{lr-regression}, \textit{sentiment}, and \textit{linpack} on GCF) or increasing (for all other functions on GCF). This can be attributed to an increase (2x) in the number of allocated clock cycles for a memory configuration of $4096$MB as compared to $2048$MB~\cite{GoogleCloudFunctionsBilling}. 
	
	
\textbf{In general, AWS lambda has a lower execution duration for most of the functions at all memory configurations as compared to GCF}: We can observe from the Figure~\ref{fig:execution_time_no_limit} that, for most of the functions except the three compute-intensive functions at a lower memory configuration  ($256MB$ and $512MB$), AWS lambda process function events faster than the GCF. For example, for \textit{dd} microbenchmark at $256MB$ configuration, AWS lambda takes $5.2x$ times less than the GCF at the same memory allocation, and even it can process faster than GCF allocated with $4GB$. \textit{nodeinfo}, \textit{sentiment}, \textit{linpack}, and \textit{web-endpoint} took almost the same amount across different memory configurations and FaaS platforms.

\subsection{Memory effect on function's concurrent instances}
\label{sec:active_instances_effect_qos}
\begin{figure}[t]
    \centering
    \includegraphics[width=1\linewidth]{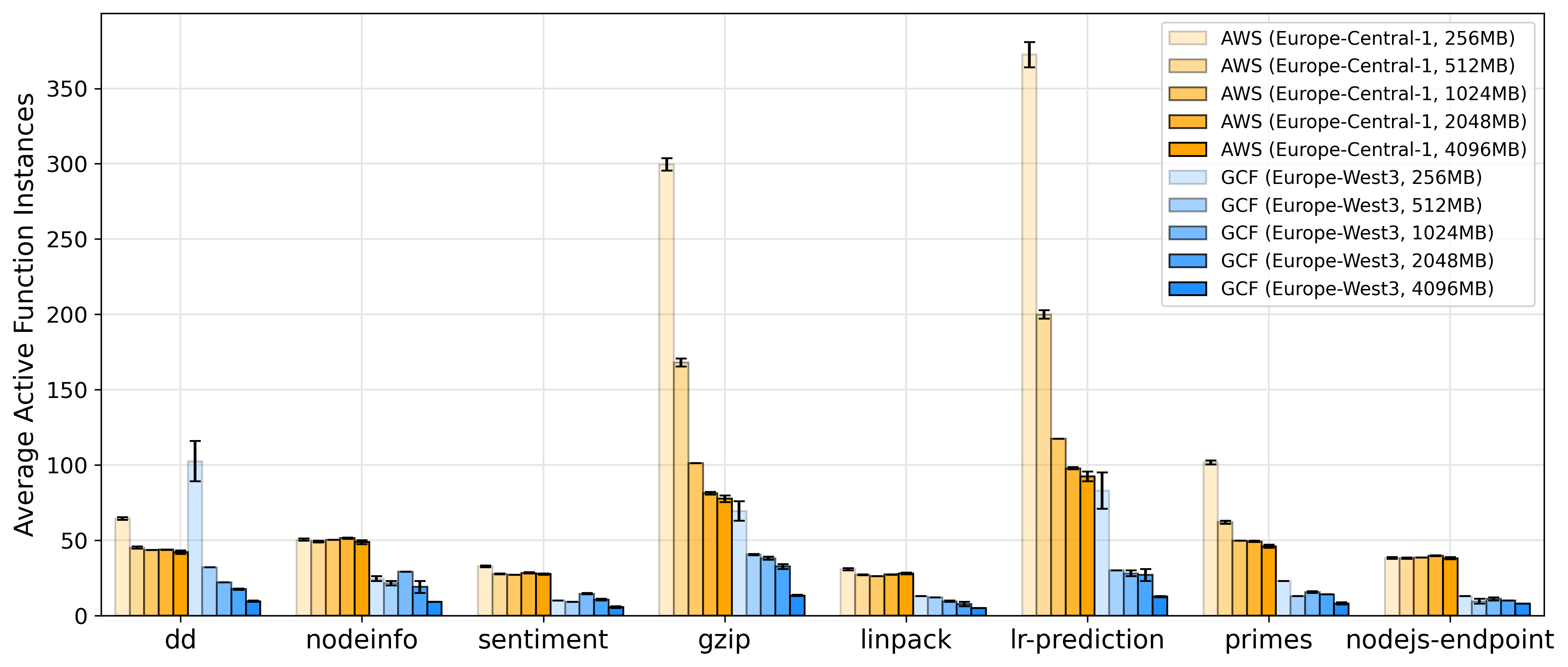}
    \caption{\textit{concurrent\_instances} of the sandboxed functions for handling the load of $50$ RPS with five different memory configurations and no limit on the \textit{function\_concurrency}. }
    \label{fig:active_instances_no_limit}
\end{figure}

Figure~\ref{fig:active_instances_no_limit} shows the \textit{concurrent\_instances} per function in the serverless application when it is load tested with a load of $50$ invocations per second on GCF and AWS lambda platform for five different memory configurations. We observe the following: 

\begin{figure*}[t]
    \centering
    \includegraphics[width=0.95\linewidth]{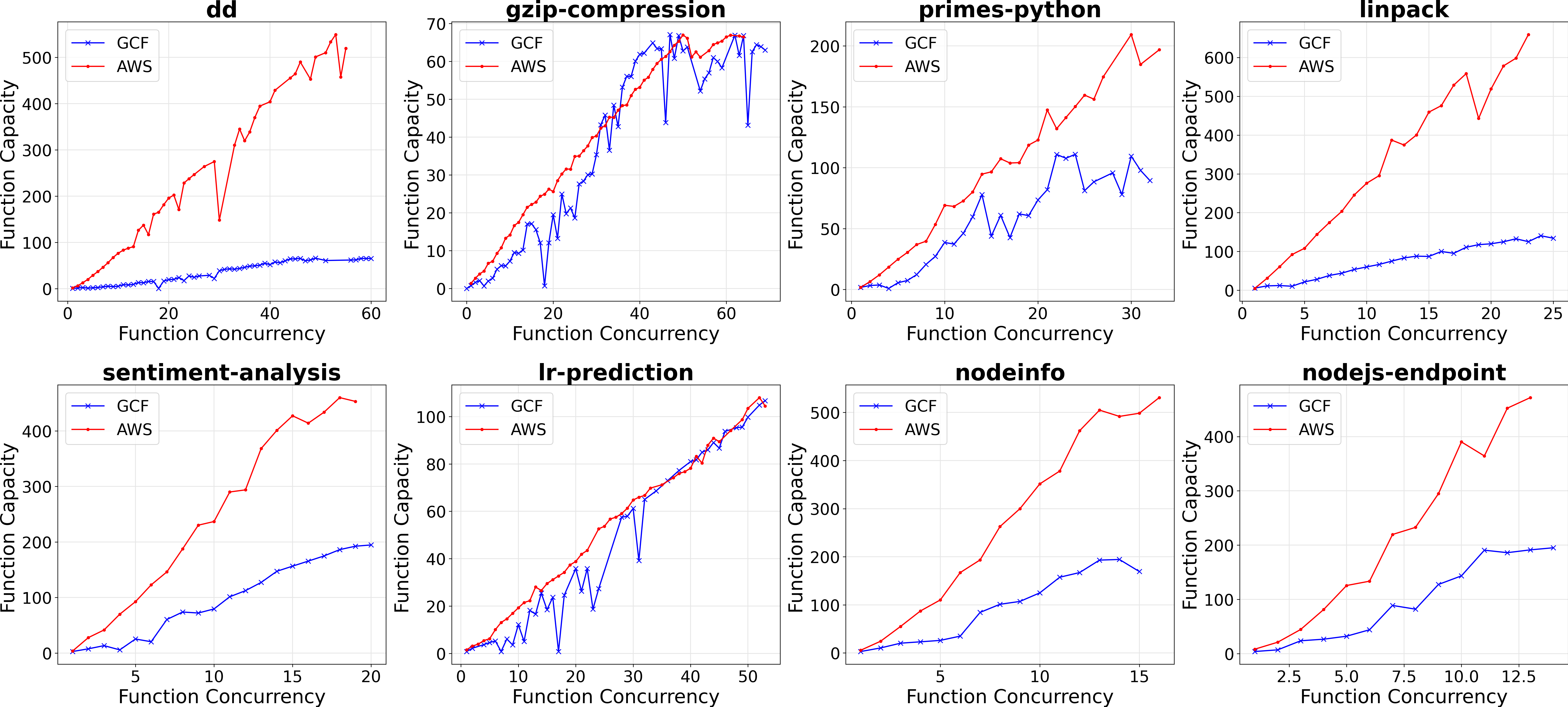}
    \caption{Function Capacity of the functions when deployed on the two FaaS platforms for different \textit{function\_concurrency} with memory configuration fixed to 256MB. }
    \label{fig:fn_capacity_actual}
\end{figure*}

\textbf{AWS lambda creates more \textit{concurrent\_instances} as compared to GCF}: From Figure~\ref{fig:active_instances_no_limit}, we can see that, for most of the functions and across different memory configurations, AWS creates a higher number of \textit{concurrent\_instances} as compared to GCF for handling the same amount of load.
	
\textbf{Decrease in number of \textit{concurrent\_instances} with the increase in memory configuration}: As the memory is increased for each function, the number of functions \textit{concurrent\_instances} for both the FaaS platforms either remain constant or has decreased. This trend can be attributed to the fact that a higher resource instance can serve the requests faster and hence can process more requests per unit time. Therefore, fewer instances are required to handle the same amount of load when allocated with lower memory configurations. 
	
\textbf{Slow-processing functions are scaled to higher number of \textit{concurrent\_instances} to match up with the fast-processing functions }:  From Figure~\ref{fig:execution_time_no_limit}, we can observe that, \textit{lr-regression} and \textit{gzip} have the highest function \textit{execution\_time} as compared to the other functions and when observing the number of \textit{concurrent\_instances} for those two functions in Figure~\ref{fig:active_instances_no_limit}, we can see that they are the highest. This concludes that the compute-intensive (slow-processing) functions require higher scaling to match up the other fast-processing functions to handle the same amount of load. Such visualization can also be used to understand the bottleneck function in the serverless application; for example, in our case, it is \textit{lr-regression}.

\subsection{Effect of function concurrency on the FC}
\label{sec:conc_effect_capacity}

Figure~\ref{fig:fn_capacity_actual} shows the actual capacity measurements for the two FaaS platforms for different \textit{function\_concurrency} configurations with memory configuration fixed to 256MB for all the functions. The capacities depicted are the average of the five runs for both FaaS platforms. In general, it can be inferred that for most of the functions, FCs vary linearly with the \textit{function\_concurrency} for both the FaaS platforms. Also, a single instance of AWS lambda can process a higher number of requests than the single instance on GCF. However, for the two compute-intensive functions namely: \textit{gzip-compression}, \textit{ml-lr-prediction} we see a similar FCs for both the platforms at different \textit{function\_concurrency}. In case of GCF, for simple web-based functions : \textit{sentiment-analysis}, \textit{nodeinfo}, and \textit{nodejs-endpoint} the linear increasing slope is not constant. From Figure~\ref{fig:fn_capacity_actual}, one can see that, for the three FaaS functions the linear slope changed after \textit{function\_concurrency} of \texttt{6}. This means that, after the \textit{function\_concurrency} of 6, each instance can process more number of requests as compared to the instance used when the \textit{function\_concurrency} is less than 6. In general, the trend is linear for all other functions and both FaaS platforms. However, they are not exactly following the ideal lines. Therefore, one needs modeling approaches for the estimation of FCs for both the FaaS platforms. 

\subsection{Function Capacity estimation}
\label{sec:capacity_prediction}

On analyzing the impact of varying memory configurations on the performance of the different FaaS functions, we use the metrics <concurrent\_instances, execution\_duration, allocated\_memory, memory\_usage, function\_concurrency> obtained from the collected load test data as input parameters for the different models (\S\ref{sec:modeler}). For a given set of input parameters, all models predict \texttt{<function invocations>} which is equivalent to the FC. 



Table~\ref{tab:modelaccuracy} shows the comparison of the accuracy results for estimated FCs for the different modeling approaches (\S\ref{sec:modeler}) with the best ones underlined for both the FaaS platforms. For determining the accuracy of the formulated models, we use the $R^2$ score~\cite{r2score}. 


\begin{table*}[t]
\caption{ Comparison of accuracy results ($R^2$ score) for estimated FCs for the different modeling approaches.}
  \label{tab:modelaccuracy}
  \centering
  \begin{tabular}{c|cc|cc|cc|cc|cc}
    \hline
    
     \bfseries Function      & \multicolumn{2}{c}{\textbf{LR}}  & \multicolumn{2}{c}{\textbf{PLR}} & \multicolumn{2}{c}{\textbf{RR}} & \multicolumn{2}{c}{\textbf{RFR}} & \multicolumn{2}{c}{\textbf{DNN}}\\
     \hline
     &\bfseries GCF& \bfseries AWS & \bfseries GCF& \bfseries AWS& \bfseries GCF& \bfseries AWS& \bfseries GCF& \bfseries AWS& \bfseries GCF & \bfseries AWS\\
    \hline
    \bfseries dd   & 81.9    &98.1&   88.27 & 97.7 & 82.2 &  98.0  &88.5&  98.1 & \underline{91.1} & \underline{98.2}\\
    \bfseries gzip   & 83.5    &91.4&   89.8 & 94.8 & 83.6 &  91.4  &93.0&  94.6 & \underline{93.6} & \underline{94.9}\\
    \bfseries primes   & 75.8   &95.4&   78.6&  95.1& 76.4 & 95.6   &83.1&  \underline{96.7}
    & \underline{85.0} & 96.5\\
    \bfseries linpack  & 86.3   &58.7&   87.7& 75.9 & 86.4 & 76.3   &88.5& 87.2  & \underline{88.8} & \underline{92.4}\\
    \bfseries sentiment  & 65.6   &33.6&   72.9& 92.9 & 52.2 &   24.9  &\underline{76.0}& \underline{97.4}  & 74.4 & 96.2\\
    \bfseries lr-pred. & 90.9  &99.4&   93.0& 99.5 & 90.7 & 99.4   &\underline{95.4}& 98.7  & 94.7 & \underline{99.5}\\
    \bfseries nodeinfo  & 80.6  & 87.5&   88.4& 87.6 & 79.6 & 87.9   &89.6& \underline{88.2}  & \underline{90.2} & 87.6\\
    \bfseries endpoint  & 77.2  & 36.5&   80.8& 67.9 & 76.3 & 35.7   &\underline{82.8}& \underline{81.0}  & 77.8 & 80.2\\
    \hline
  \end{tabular}
\end{table*}


In general, it was found that the accuracy measurements for Function Capacity estimation for AWS lambda are higher than the GCF for most of the FaaS functions since AWS lambda exhibits more linear behavior as compared to GCF ( Figure~\ref{fig:fn_capacity_actual}). \textbf{Linear Regression (LR)} leads to best results when the parameters are linearly correlated to the FC. For most FaaS functions, the parameters are linearly correlated to FC, leading to an accuracy value greater than $80$\%. For both the FaaS platforms, the accuracy for the \textit{nodejs-endpoint}, and \textit{sentiment-analysis} FaaS functions is comparatively less as compared to other FaaS functions since most parameters in them are non-linearly correlated with FC.  \textbf{Polynomial Linear Regression (PLR)} leads to highly accurate results for most of the function types due to its ability to model non-linear relations among the parameters.  With PLR, the estimated FC accuracies are higher than the linear regression, which is attributed to most parameters being non-linearly correlated. \textbf{Ridge Regression (RR)} produced approximately the same results as that of linear regression and worked well for certain function types. On the other hand, \textbf{Random Forest Regression (RFR)} can provide the best results among the statistical approaches.

\begin{figure}[t]
    \centering
    \includegraphics[width=0.7\linewidth]{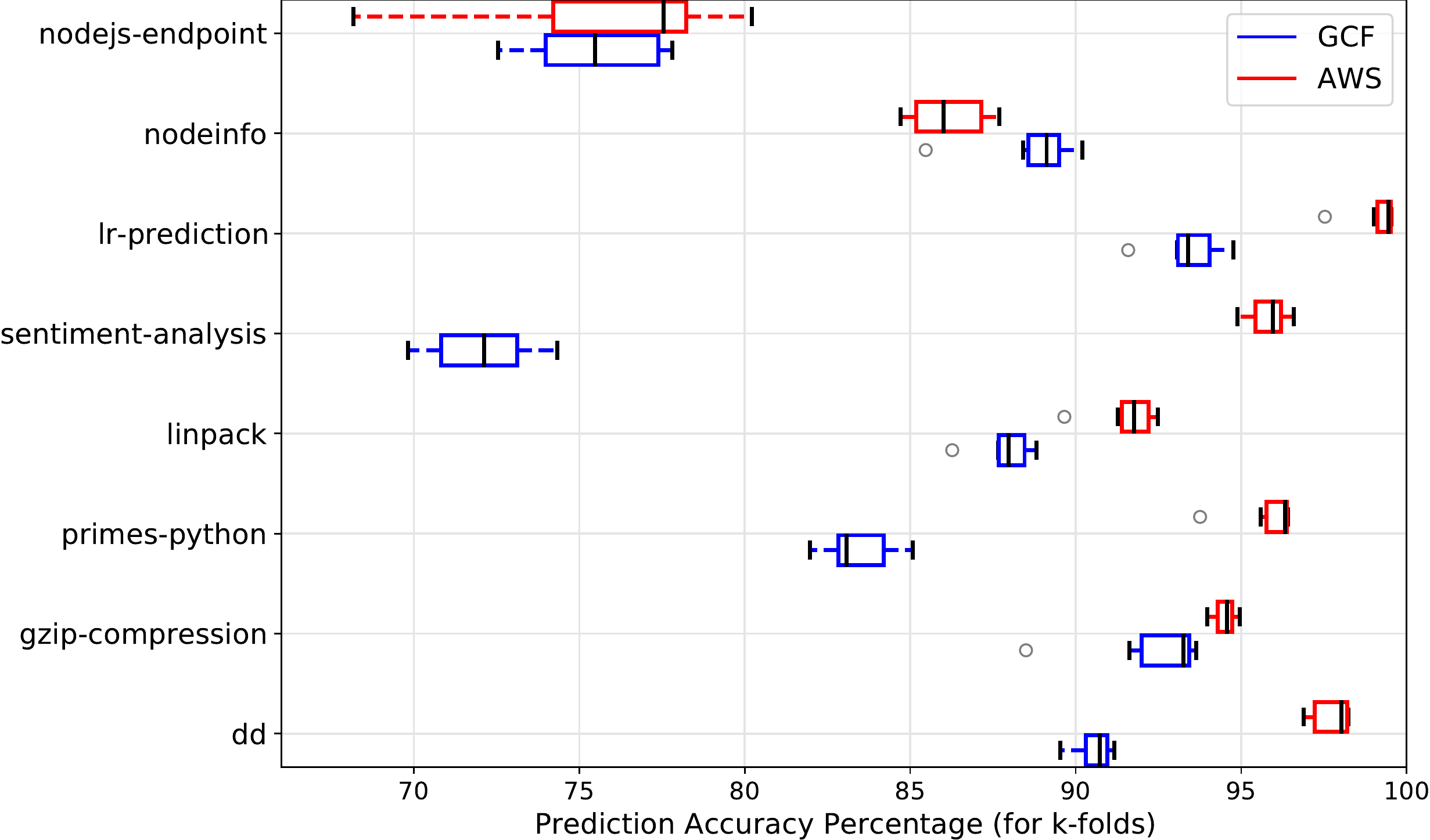}
    \caption{Box plot showing the prediction accuracy on the test data across k-folds using DNN model for both FaaS platforms.} 
    \label{fig:prediction}
\end{figure}

The \textbf{Deep Neural Network(DNN)} method outperformed all the statistical approaches for most of the FaaS functions since it is capable of modeling the linear and non-linear correlation between the parameters. For most function types, the FC estimation accuracy is approximately above $75\%$. In Figure~\ref{fig:prediction}, we show the prediction accuracy percentage for k-folds in the case of DNN on the test data using the box plot for both FaaS platforms, and all the FaaS functions. 

\section{Related Work}
\label{related_work}

FaaS is priced as per the pay-as-you-go model, where they charge the number of function invocations and memory. The majority of the prior work is based on optimizing the scalability, cost, execution time, integration support, and the constraints associated with FaaS services such as AWS Lambda, GCF, and Azure Functions (AF) provided by public cloud  providers~\cite{grogananalysis, closer20, fdn, 9546403, 10.1007/978-3-030-72369-9_8}. However, there is no prior research for quantifying and estimating the FCs based on the different deployment configurations using GCF and AWS Lambda. We present prior work from two aspects:  

\subsection{Performance impact of the underlying system}
FaaSProfiler~\cite{shahrad2019architectural} is the first to take a bottom-up approach in analyzing the architectural implication to unwrap the server-level overheads in the FaaS model. They analyzed the difference between native and in-FaaS function execution and calculated the additional server-level overheads like computational overheads, memory consumption, bandwidth usage, and management overheads like orchestration, queuing, scheduling, and power consumed. Furthermore, Lee et al.~\cite{lee2018evaluation} compared the performance of various serverless computing environments offered by public cloud providers by showcasing the results of throughput, network bandwidth, file I/O and compute performance on concurrent function invocations. Wang et al.~\cite{wang2018peeking} performed an in-depth study of resource management and performance isolation with three popular serverless computing providers: AWS Lambda, Azure Functions, and GCF. Their analysis demonstrates a reasonable difference in performance between the FaaS platforms. Moreover, the deployed function instances are associated with VMs having a variety of configurations with different underlying host micro-architectures. In this work, we showcase a strong correlation between the function's performance and the resources allocated to it based on the different configurations.

Additionally, Figiela et al.~\cite{figiela2018performance} developed a cloud function benchmarking framework. Compute-intensive functions were deployed in major cloud providers' FaaS platforms. The authors observe variation in response time duration based on the different underlying hardware and runtime environment. These observations encouraged us to proceed with estimating the FCs of functions when deployed with different deployment configurations. 



\subsection{Performance modeling of the FaaS application}
Pawlik et al.~\cite{pawlik2018performance} state that to assess the feasibility of running an application on the FaaS platform, we have to determine the SLO of the application. This can be achieved by constructing a reliable performance model capable of analyzing a function performance, which requires knowledge about the performance of the infrastructure. Cloud service providers abstract details such as the number of cores, memory available, and network I/O capacity in the underlying hardware, usually limiting the available information to function time limit, maximum memory. The allocated memory also affects the provisioned CPU quota~\cite{perfcompute}. In our previous work~\cite{jindal2019performance}, we developed a tool for estimating the capacity of a microservice application when it is sandboxed. We followed the similar approach in this work for the capacities estimation of the FaaS functions.

\section{Conclusion and Future Work}
\label{conclusion}
In this paper, we demonstrated the impact of various configuration parameters on the Function Capacity of the two FaaS platforms(AWS Lambda and GCF). The methodology and the tool \textit{FnCapacitor} introduced in this paper aim to solve the problem of estimating the Function Capacity at a specific deployment configuration. In summary, \textit{FnCapacitor} can be used in two ways: 
\begin{itemize}
    \item By application developers in an offline manner for estimating the FCs of FaaS functions within their application for different deployment configurations. Based on the estimated FCs and the requirements, the developer can deploy the functions with the right configurations.
     \item By application developers in an online manner, the tool collects the monitoring data of the already existing FaaS functions. It automatically builds the models in the background without additional load testing. The built models can then be used to update the deployment configurations of the functions depending on the required SLOs. 
\end{itemize}


In the future, we plan to extend \textit{FnCapacitor} with other public serverless compute providers and to open source FaaS platforms. Creating a function scheduling framework based on the estimated FCs, the functions can be created with the suitable deployment configurations is another future direction. 

\begin{acks}
This work was supported by the funding of the German Federal Ministry of Education and Research (BMBF) in the scope of the Software Campus program. Google Cloud credits in this work were provided by the \textit{Google Cloud Research Credits} program with the award number \grantnum{NH93G06K20KDXH9U}{NH93G06K20KDXH9U}.

\end{acks}
\bibliographystyle{ACM-Reference-Format}
\bibliography{bib}

\end{document}